\documentclass[a4paper]{jpconf}
\usepackage{graphicx}
\usepackage{amssymb}
\usepackage{amsmath}
\usepackage{bm}
\usepackage{color}

\newcommand{\prl}{Ref.~\cite{Eckstein2010}}
\newcommand{\expval}[1]{\langle #1 \rangle}
\renewcommand{\dh}{dh}
\newcommand{\jd}{\Gamma_\text{\dh}}
\newcommand{\fth}{F_\text{th}}
\newcommand{\hopp}{{t^*}}

\begin{document}
\title{Dielectric breakdown of Mott insulators -- doublon production and doublon heating}

\author{Martin Eckstein$^1$ and Philipp Werner$^2$}

\address{$^1$ Max Planck Research Department for Structural Dynamics, University of Hamburg, CFEL, Hamburg, Germany}
\address{$^2$ Department of Physics, University of Fribourg, 1700 Fribourg, Switzerland}

\ead{martin.eckstein@mpds.cfel.de}


\begin{abstract}
Using dynamical mean-field theory and the non-crossing approximation as impurity solver, we study the response of a 
Mott insulator to strong dc electric fields. The breakdown of the Mott insulating state is triggered by field-induced  creation 
of doublon-hole pairs. In a previous investigation, Ref.~\cite{Eckstein2010}, it was found that the system approaches a 
long-lived quasi-steady state in which the current is time-independent although the number of carriers constantly increases. 
Here we investigate and clarify the nature of this state, which exists only because thermalization is slow in the Hubbard model 
at strong coupling. The current is time-independent because doublons and holes have an infinite temperature distribution. 
Evidence for this fact is obtained from spectral functions and by comparing the electric current with the field-induced 
doublon-hole creation rate. Implications to real experiments, in systems with energy dissipation, are discussed.
\end{abstract}


\section{Introduction}
\vspace{2mm}

The study of nonequilibrium phenomena in complex strongly correlated systems may reveal unexpected physics, 
which could eventually lead to new ways of tuning and controlling material properties on ultrafast timescales. 
The Mott insulator, a system with partially filled bands in which electrons are localized due to the strong 
Coulomb repulsion, is an ideal example of a correlated many-body state, and Mott insulators are
common among transition metal oxides or organic charge-transfer salts \cite{Imada1998}. 
A possible way to drive these systems out of equilibrium is to expose them to a strong (dc) electric field $F$. 
The resulting dc response, which is non-perturbative in the field and thus cannot be explained by response 
functions of the equilibrium state, is known as  the dielectric breakdown. A considerable amount of work in 
the previous few years has been devoted to understand this conceptually simple, and yet theoretically 
challenging phenomenon.  Experimentally, nonlinear transport in correlated insulators  has been studied 
in both oxides \cite{Tokura1988,Guenon2012} and in organic materials \cite{Inagaki2004,Sawano2005}.  
One observes a strong non-linearity in the current-field ($j$-$F$) characteristics, with a negative differential 
resistivity between weak current and  large current regimes. The ``strong-field physics'' of correlated materials 
can also be addressed in experiments with ultra-cold atoms in optical lattices, e.g., as an 
elementary probe of the Mott phase~\cite{Greiner2002}.

In an intuitive picture, the Mott insulating ground state decays through the production of doublon-hole pairs
in the presence of an electric  field: In a scalar potential gauge, the field implies a potential energy difference $F$ 
between two neighboring lattice sites in the field-direction (choosing units with lattice spacing $a=1$ and electron 
charge $e=1$). In the half-filled Mott insulator, with unit occupancy at every site, an electron can thus gain the 
energy $U$ from the field which is needed to form a doublon (doubly occupied site) and leave behind a hole, 
by tunneling over a distance $\ell_U=U/F$. In this picture, the tunneling process alone contributes a current of the 
order
\begin{equation}
\label{doublon-current}
\jd (t)  = \ell_U \dot d(t),
\end{equation}
where $\dot d$ is the rate of doublon-hole (\dh) pair production. In addition, doublons and holes are expected to 
carry a current after they have been created. We will refer to $\jd$ as the \dh-creation current. Although its definition 
may seem ad-hoc here, $\jd$ will turn out to be very useful in the analysis of the data below.

For $F \ll 1$ the tunneling is over many lattice sites and thus occurs at an exponentially small rate.
An exponential scaling with a threshold field $\fth$ for the electric current,
\begin{equation}
\label{fth}
j \propto F \exp
\Big(
-\frac{\fth}{F}
\Big),
\end{equation}
and for observables related to the doublon-production rate (with different powers of $F$ in the pre-factor) is indeed 
found quite generically, for the fermionic Hubbard model in one dimension \cite{Oka2003a,Oka2005,Meisner2010,
Kirino2010,Oka2010a,Oka2012,Lenarcic2012}, in infinite dimensions \cite{Eckstein2010}, and also for the bosonic 
Mott insulator \cite{Queisser2012}. 
A lot of work aimed at understanding the dependence of $\fth$ on the charge gap $\Delta_c$.
Motivated by the dielectric breakdown of band insulators \cite{Oka2005}, which can be explained by the Landau-Zener 
mechanism \cite{LZ}, Oka {\it et al.} proposed a description of the Mott breakdown in the one-dimensional Hubbard model 
in terms of Landau-Zener tunneling between many-body eigenstates \cite{Oka2003a,Oka2005,Oka2010a,Oka2012,Oka2005b}. 
Both numerical work \cite{Oka2003a,Oka2005} using exact diagonalization and density-matrix renormalization 
group (DMRG), and an analytical approach involving the ground state and the first excited state of the Bethe ansatz 
solution \cite{Oka2010a,Oka2012} then suggest a relation $\fth \sim \Delta_c^2$ for small $U$. Another analytical approach 
is the solution for the case of one spin-$\downarrow$ electron in a spin-$\uparrow$-polarized background, for which one finds 
$\fth\propto\Delta_c^{3/2}$  \cite{Lenarcic2012}. DMRG results suggest that this behavior holds down to smaller 
polarization (the Mott insulator being the unpolarized case). Furthermore, if the voltage drop is applied over one 
lattice site rather than linearly, one has $\fth\propto\Delta_c$ \cite{Kirino2010}. In the infinite-dimensional case 
\cite{Eckstein2010}, $\fth$ clearly increases with $\Delta_c$, but the region close to the metal insulator transition 
($\Delta_c\to0$) has so far not been studied.

In the present paper we do not focus on the value of $\fth$ close to the metal-insulator transition,
but we want to clarify another fundamental question that arises in connection with studies of the dielectric 
breakdown:  Most of the theories mentioned above involve isolated systems. After the field is switched on, one 
observes the emergence of a quasi-steady state in which doublons are produced at a more or less  time-independent 
rate, and dc properties of the system are obtained from this quasi-steady state. When the electric current is computed 
independently, one finds that $j$ is also time-independent  \cite{Eckstein2010,Meisner2010,Mikelsons2012}. 
This seems quite peculiar if doublons and holes, whose number is constantly increasing,  are interpreted as 
``charge carriers''. The main purpose of the present paper is to clarify the nature of this quasi-steady state: 
We demonstrate that its existence is related to a failure of the quasi-equilibrium description, i.e., a lack of 
thermalization (Sec.~\ref{sec-quasi-failure}). We furthermore clarify the relation of  the electric current and the 
doublon production rate, and provide an explanation why the doublons and holes which are generated by the 
field appear to be immobile and do not contribute to the current (Sec.~\ref{sec-temp}). This also leads to 
a better understanding of how temperature (Sec.~\ref{sec-temp}) or generic energy dissipation mechanisms 
(Sec.~\ref{sec-diss}) influence the dielectric breakdown, which is essential for understanding (and designing) 
experiments that can probe the non-perturbative effects of Eq.~(\ref{fth}) (Sec.~\ref{experiments}). Our analysis builds on the results
of Ref.~\cite{Eckstein2010}, in which the dielectric breakdown was studied in the limit of large dimensions 
\cite{Metzner1989a}, using dynamical mean-field theory (DMFT) \cite{Georges1996}.


\section{Model and methods}
\label{methods}
\vspace{2mm}

Throughout this paper we study the paramagnetic Mott insulating phase in the half-filled Hubbard model on a $d$-dimensional 
cubic lattice,
\begin{equation}
\label{hubbard}
H
=
\sum_{\langle ij \rangle\sigma} 
V_{ij}(t) \,
c_{i\sigma}^\dagger c_{j\sigma}
+
U\sum_i (n_{i\uparrow}-\tfrac12)(n_{i\downarrow}-\tfrac12).
\end{equation}
Here $c_{i\sigma}^\dagger$ ($c_{i\sigma}$) is the creation (annihilation) operator for an electron with spin $\sigma$ at lattice site 
${\bm R}_i$, $U$ denotes the Coulomb repulsion, and $V_{ij}$ is the matrix element for hopping between nearest neighbor sites $i$ and $j$. We 
initially prepare the system in thermal equilibrium at temperature $T=1/\beta$, and apply a homogeneous electric field 
${\bm F}(t)$ for time $t>0$. For convenience, $\bm F$ is always pointing along the body diagonal $\hat{\bm\eta}=(1\ldots1)^t$,
which simplifies the DMFT self-consistency (see below) \cite{Turkowski2005}.
The field is turned on to a value $F$ within a switching time $t_0$ with a given ramp profile $r(x)$,
\begin{align}
\bm F(t) 
&= \hat{\bm \eta} \,F\, r(t/t_0),
\\
r(x)
&= 
\left\{
\begin{array}{cl}
\tfrac12-\tfrac34\cos(\pi x)+\tfrac14 \cos(\pi x)^3 & \text{~~for~} 0\le x\le 1\\[1mm]
1 & \text{~~for~} x \ge .
\end{array}
\right.
\end{align}
The smooth turn-on of $F$ damps transient currents, but does not influence the long-time behavior \cite{Eckstein2010}. 
To incorporate the field into Eq.~(\ref{hubbard}) we use a gauge with pure vector potential ${\bm A}(t)$, i.e., 
${\bm F}(t)=-\partial_t {\bm A(t)}/c$,  for which 
the 
hopping matrix elements acquire a Peierls phase, 
\begin{equation}
V_{ij}(t)=V_{ij}^0 e^{ i e ({\bm R}_j-{\bm R}_i){\bm A}(t)/\hbar c }.
\end{equation}
(For a discussion of various gauges within a tight-binding model, see Ref.~\cite{Davies1988}.) 
In the limit $d=\infty$ \cite{Metzner1989a}, with rescaled hopping $ V_{ij}^0=\hopp/2\sqrt{d}$,  the problem can be solved 
exactly using DMFT \cite{Georges1996} in its nonequilibrium (Keldysh) formulation \cite{Schmidt2002,Freericks2006a}. 
The rescaled nearest neighbor hopping $\hopp$ is used as the unit of energy, such that the density of states 
is given by $\rho(\epsilon) \propto \exp(-\epsilon^2)$. Time and field are measured in units of $\hbar/\hopp$ and 
$\hopp/ea$, respectively ($\hbar=1$, $e=1$, $a=1$). 

The DMFT single-site problem is solved by means of the self-consistent hybridization expansion 
\cite{Eckstein2010b}. The entire set of equations for the geometry considered here has been discussed 
in Ref.~\cite{Eckstein2011}, and details of the numerical implementation of the Keldysh equations are given in 
Ref.~\cite{Eckstein2010a}. In the present work we use the lowest order of the strong-coupling solver,
or non-crossing approximation (NCA) \cite{NCA}, which allows us to address certain issues that require long 
simulation times.  In order to assess the validity of this approximation, we compare these results in 
Sec.~\ref{sec-current} to the results of {\prl}, which were obtained using the second order, 
or one-crossing approximation (OCA) \cite{Pruschke1989}.
 
\subsection*{Observables}
\vspace{1mm}

From the DMFT solution we directly evaluate  \cite{Eckstein2010a}
expectation values of the double occupancy
$
d=\big\langle
\frac{1}{L} \sum_i  \hat  n_{i\uparrow} \hat n_{i\downarrow} 
\big\rangle
$
and the kinetic energy
$
E_\text{kin}
=
\big\langle
\frac{1}{L} 
\sum_{{\bm k}\sigma} \epsilon_{{\bm k} + {\bm A(t)}} 
\hat n_{{\bm k}\sigma}
\big\rangle
$,
where the band energy 
$\epsilon_{\bm k}$
is the Fourier transform of $V_{ij}^0$.
The current is given by
$
{\bm j}
=
\big\langle
\frac{1}{L} 
\sum_{{\bm k}\sigma} {\bm v}_{{\bm k} + {\bm A(t)}} 
\hat  n_{{\bm k}\sigma}
\big\rangle
$,
with the band velocity ${\bm v}_{\bm k}=\partial_{\bm k} \epsilon_{\bm k}$.
One can then verify the general fact that the total internal energy
\begin{equation}
E_\text{tot} =
E_\text{kin}(t)
+
U d(t)
\end{equation}
changes according to the equation
\begin{equation}
\label{energy-increase}
\dot E_\text{tot}(t) = {\bm j}(t) {\bm F}(t).
\end{equation} 
Because the self-consistent strong-coupling solver is conserving in the sense of 
Kadanoff and Baym, Eq.~(\ref{energy-increase}) is satisfied also for the approximate solution.

In addition to the static observables, we define a spectral function from the Fourier transform of the retarded local Green 
function $G^\text{ret}(t,t')= -i \Theta(t-t')\langle \{c(t),c^\dagger(t')\}\rangle$,
\begin{equation}
\label{aret}
A(\omega,t)
=
-\frac{1}{\pi}
\text{Im}
\int_0^{s_\text{max}}
\!ds\,
e^{i\omega s}\,
G^\text{ret}(t+s,t).
\end{equation}
For a general non-equilibrium situation, $A(\omega,t)$ is not necessarily positive. However, when the $t$-dependence of 
$A(\omega,t)$ can be neglected, $A(\omega,t)$ is positive, and it can be 
related to photoemission and inverse photoemission spectra in the usual way \cite{PES}. In a quasi-steady
state, as we will encounter below, these properties remain as long as the $t$-dependence of $A(\omega,t)$ is very slow compared to the 
inverse of the scale $\Delta\omega$ on which $A(\omega,t)$ changes as a function of $\omega$. Technically,
a finite cutoff $s_\text{max}$ limits the frequency resolution. In analogy to Eq.~(\ref{aret}) we will also look at the 
density of occupied states (corresponding to the photoemission spectrum),
\begin{equation}
\label{ales}
A^<(\omega,t)
=
\frac{1}{\pi}
\text{Im}
\int_0^{s_\text{max}}
\!ds\,
e^{i\omega s}\,
G^<(t+s,t),
\end{equation}
which is the Fourier transform of the lesser Green function $G^<(t,t')= i \langle  c^\dagger(t') c(t)\rangle$.
The occupation function 
\begin{equation}
N(\omega,t)= A^<(\omega,t)/A(\omega,t)
\end{equation} 
gives the Fermi function $f(\omega)$ in equilibrium.


%
\begin{figure}[t]
\centerline{\includegraphics[width=16cm,clip=true]{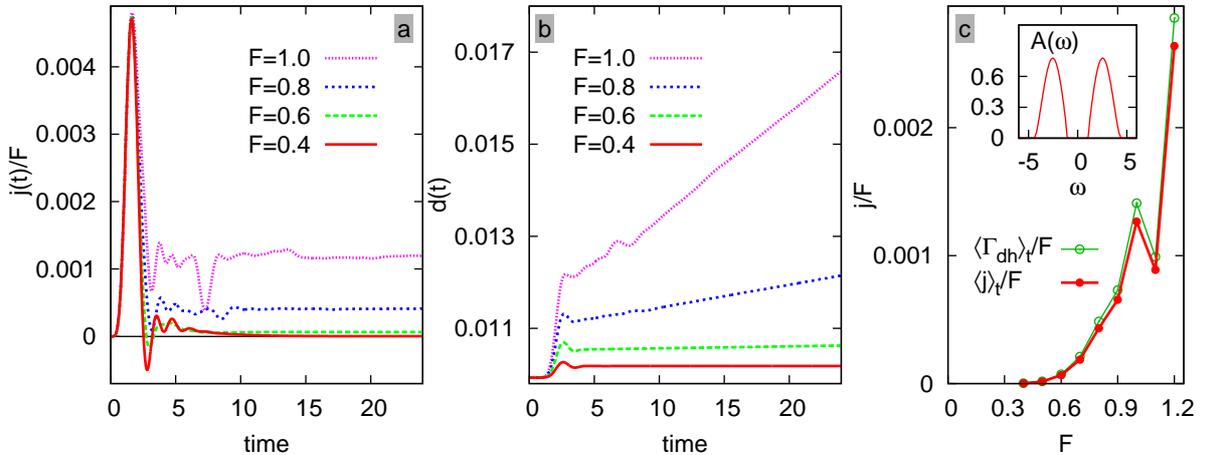}}
\caption{\label{fig-current_U5b5}
(a) Time-dependent current $j(t)$ for $U=5$ and $\beta=5$ and various electric fields $F$. (b) Double occupancy 
for the same parameters. (c) Current $\expval{j}_t$, averaged over times $40<t<50$. The curve  $\expval{\jd}_t$ 
(open symbols) shows corresponding data for the doublon current Eq.~(\ref{doublon-current}), see Sec.~\ref{sec-temp}. 
{\it Inset:} Spectral function for the equilibrium state at $U=5$ and $\beta=5$.}
\end{figure}
\section{The dielectric breakdown current}

\vspace{2mm}
\subsection{The current}
\label{sec-current}
\vspace{1mm}

In this section we briefly recapitulate the main results of {\prl}. We use the lowest order of the strong-coupling expansion 
(NCA) to compute several quantities related to the dielectric breakdown, and compare the outcome to OCA results from 
{\prl}. 
Figure \ref{fig-current_U5b5}a shows the current $j(t)$ for various electric fields $F$ as a function of time, for
$U=5$ and $\beta=5$. The spectral function for these parameters, with a well-developed gap at the Fermi energy, 
identifies the initial state as a Mott insulator (inset of Fig.~\ref{fig-current_U5b5}c). The temperature $1/\beta=0.2$ 
is still too low to allow for a sizable number of thermally excited carriers, such that the linear dc conductivity is indistinguishable 
from zero on the scale of  the plot. Nevertheless, the value of $j(t)$ at long times is nonzero and it strongly increases 
with $F$. (The large signal for $t \lesssim 3$, which is linear in $F$, is the current 
associated with the polarization of the insulator.) The highly nonlinear $j$-$F$ characteristics of the Mott insulator 
is revealed by plotting the long-time average $\expval{j}_t$ of the current against $F$ (Fig.~\ref{fig-current_U5b5}c). 
The threshold-like increase of $\expval{j}_t/F$ around $F=0.6$ is the hallmark of the dielectric breakdown of the 
Mott insulator.

\begin{figure}[t]
\centerline{\includegraphics[width=16cm,clip=true]{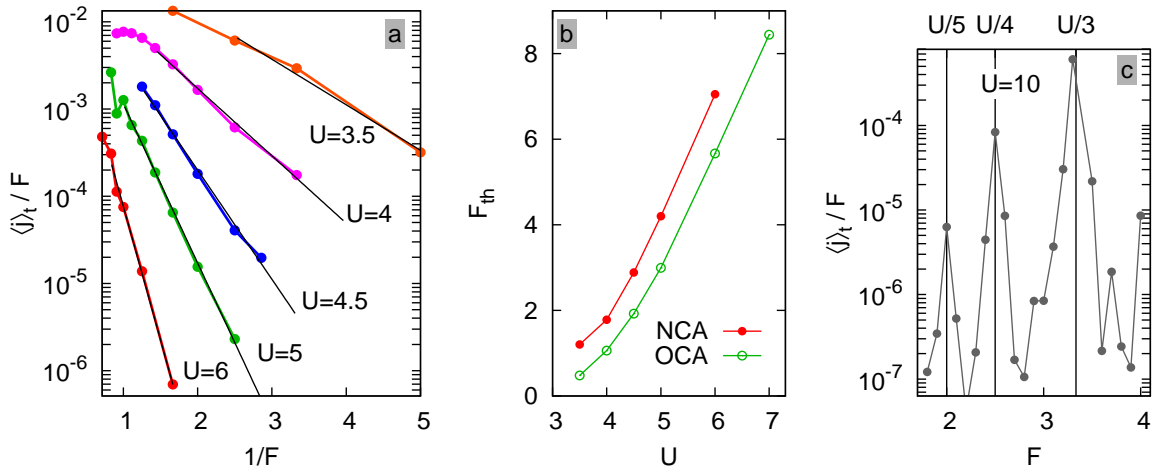}}
\caption{\label{fig-fth}
(a)
Current $\expval{j}_t/F$ (averaged over times $40<t<50$)  on a logarithmic scale as a function of $1/F$ 
for various values of $U$ ($\beta=5$). In this plot, black solid lines correspond to fits with Eq.~(\ref{fth}), 
in the intervals covered by the lines.
(b)
The value $\fth$, obtained from the linear fits (filled symbols). Open symbols show corresponding results 
from {\prl}, which were obtained within the OCA.
(c)
Current $\expval{j}_t/F$ (averaged over times $20<t<30$)  on a logarithmic scale,
for large $U$ and $F$ (see discussion in Sec.~\ref{sec-spectrum}). Vertical lines indicate resonances at 
integer $U/F$.}
\end{figure}

Following \prl, we fit the $j$-$F$ curves with Eq.~(\ref{fth}) in order to determine the threshold field $\fth$ (Fig.~\ref{fig-fth}).
Note that Eq.~(\ref{fth}) can only be expected to hold asymptotically for small $F$. For large $F$, the  $j$-$F$ curves
can even behave non-monotonously (see, e.g., $F>1$ for $U=5$), which will be explained in Sec.~\ref{sec-spectrum} below. 
On the other hand, the exponentially small current for $F\ll1$ cannot easily be resolved numerically. This restricts the fits with 
Eq.~(\ref{fth}) to an intermediate range of $F$, as indicated by the black solid lines in Fig.~\ref{fig-fth}a. The resulting threshold 
field $\fth$, which is shown in Fig.~\ref{fig-fth}b, decreases when $U$ is decreased from the insulating regime towards the 
metal-insulator transition (which is only a crossover at $\beta=5$). 

In Fig.~\ref{fig-fth}b, we have also included OCA data from {\prl}  for the same parameters (open symbols). One can see that 
the higher-order corrections of the OCA are clearly important to get a quantitatively correct value of $\fth$. The NCA threshold 
curve is shifted to smaller interactions compared to data from {\prl}. This behavior resembles the difference between OCA 
and NCA in equilibrium, e.g., for the location of the metal-insulator transition line. Apart from that, however, we find that 
the physics obtained with NCA throughout the insulating regime is qualitatively similar to the OCA solution (compare, e.g., 
Fig.~\ref{fig-fth} with Fig.~4c of \prl). Further discussions in this paper will thus only be based on NCA results. This allows us to 
study in detail several relaxation processes with and without dissipation, and to obtain high-resolution results for the spectral 
function. The calculation of the latter requires data at long times that would be accessible within the OCA only at large numerical 
expense.


%
\begin{figure}[t]
\centerline{\includegraphics[width=16cm,clip=true]{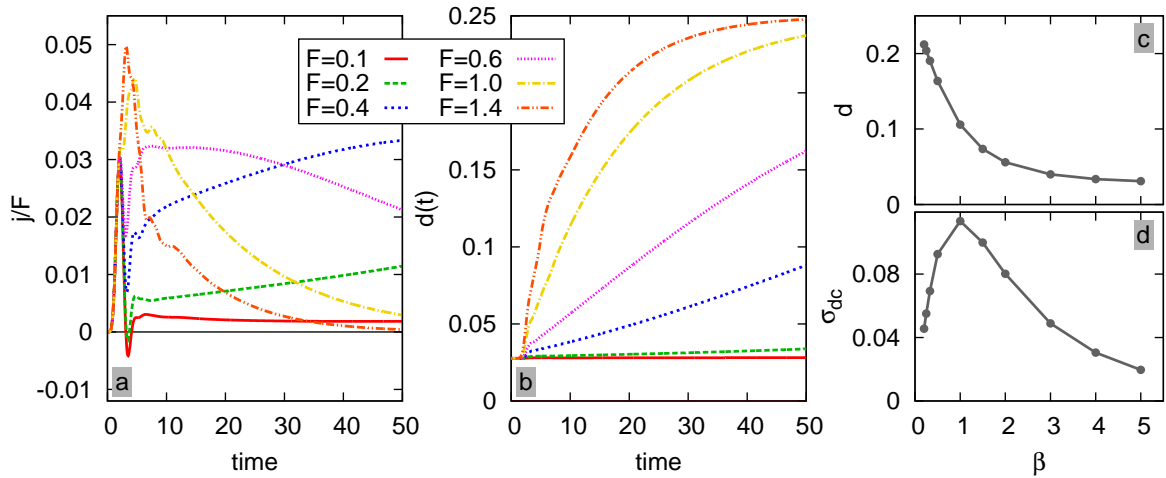}}
\caption{\label{fig-quasiequi}
(a) Time-dependent current $j(t)$ for $U=3$ and $\beta=5$ and various electric fields $F$. 
(b) Double occupancy for the same parameters. 
(c) Double occupancy $d(T)$, and (d), linear response conductivity $\sigma_\text{dc}(T)$,
in thermal equilibrium  at $U=3$ and temperature $T=1/\beta$.
}
\end{figure}

\subsection{Failure of the quasi-equilibrium description}
\label{sec-quasi-failure}
\vspace{1mm}

Although the current in Fig.~\ref{fig-current_U5b5}a becomes almost stationary at long times, the system is not in a true steady state. 
This is already clear from a very general energy consideration, because the total energy of any closed system in an external field $F$ 
must  increase at the rate given by Eq.~(\ref{energy-increase}).
Because the energy per particle is bounded from above in a single-band model, a time-independent current cannot last forever. In 
general, the  energy increase involves both kinetic and interaction energy. The double-occupancy $d(t)$, which is proportional to 
the interaction energy, is plotted in Fig.~\ref{fig-current_U5b5}b. We find that $d(t)$ indeed increases almost linearly with time, while $j(t)$ does 
not change substantially (compare also Fig.~1b of {\prl}). This behavior is observed (in a quasi-steady manner) for rather long times, 
although in principle $d$ cannot exceed the value $0.5$. As mentioned in the introduction, it is this peculiar quasi-steady state from 
which properties of the dielectric breakdown are usually inferred. In this section we investigate to what extent a simple quasi-equilibrium 
description can account for the observed behavior.  

In many situations, it is a valid assumption that a system rapidly evolves to a new equilibrium state at elevated temperature
after the energy is increased. In accordance with this, the simplest possibility to describe an isolated system in an external field
is to assume that all its properties can be obtained from an effective equilibrium state, with a time-dependent temperature 
$T_\text{eff}(t)$ which follows from the equation $C_V \dot T_\text{eff}(t)= j(t)F(t)$. In particular, the current at small fields would 
be given by 
\begin{equation}
\label{j-quasieq}
j_\text{quasi-eq.}(t) = F \sigma_\text{dc}(T_\text{eff}),
\end{equation}
where $\sigma_\text{dc}$ is the linear conductivity which becomes nonzero for $T_\text{eff}>0$. This argument was found to apply to 
various correlated systems \cite{Mierzejewski2010,Mierzejewski2011,Eckstein2011b}.  In the Hubbard model at weak-coupling, 
e.g., the quasi-equilibrium description works if $U$ exceeds some critical interaction, while for smaller interaction scattering of the particles 
is too slow to establish the equilibrium state, and the system performs long-lived Bloch oscillations \cite{Eckstein2011b}.

From the time-independence of $j(t)$ in Fig.~\ref{fig-current_U5b5}a we can already infer that a simple quasi-equilibrium description is not valid 
for the dielectric breakdown at $U=5$: Because the linear-response current in the insulator depends exponentially on temperature, a 
quasi-equilibrium state which is consistent with the increase of $d(t)$ would imply a current which is far too large. In the case of 
the Mott insulator at $U=5$ and $F=1$, e.g., the double occupancy increases by $0.006$ from $t=10$ to $t=35$ (Fig.~\ref{fig-current_U5b5}b), 
while the conductance remains at $ j/F \approx0.001$. On the other hand, in order to increase $d$ by $0.006$ {\em in equilibrium}, 
one would have to increase the temperature from $\beta=5$ ($d\approx0.010$) to $\beta=2$ ($d\approx0.016$), but at $\beta=2$, 
the linear response conductivity is already more than one order of magnitude larger than the quasi-steady value at $F=1$ 
[$\sigma_\text{dc}(\beta=2)\approx0.0155$]. 

The lack of thermalization in the Hubbard model at $U \gg \hopp $ has been encountered previously \cite{Kollath2007a,Eckstein2009a}, 
and it implies the existence of various interesting metastable states in the Hubbard model \cite{Petrosyan2007,Rosch2008,Werner2012a}.
Slow thermalization is ultimately related to the fact that the recombination time  of doublons and holes is exponentially long for $U \gg \hopp$ 
\cite{Strohmeier2010a,Sensarma2010a}, and hence kinetic energy and interaction energy cannot efficiently be redistributed. 
The long lifetime of doublons has been suggested earlier as a prerequisite for the existence of a quasi-steady current
\cite{Eckstein2010,Mikelsons2012}. Because the thermalization time $\tau_\text{th}$ of the double occupancy in the paramagnetic Mott 
insulator has recently been determined within DMFT (cf.~Fig.~2b of Ref.~\cite{Eckstein2011}, for a final temperature $\beta=2$), we 
can make the argument more quantitative in the following. For $U=5$, a value $\tau_\text{th}\approx4000$ has been found (within NCA), 
which clearly exceeds the scale of Fig.~\ref{fig-current_U5b5}. On the other hand, $\tau_\text{th}$ can become as short as a few inverse 
hoppings when $U$ is in the metal-insulator crossover regime, and it will thus be interesting to see what a more rapid thermalization 
implies for the behavior of the current. Figure~\ref{fig-quasiequi} shows $j(t)$ for $U=3$ (where Ref.~\cite{Eckstein2011} gives 
$\tau_\text{th}\approx 40$ at $\beta=2$.) In this case the system does indeed no longer establish a time-independent current. 
Instead, $j(t)$ increases with time as long as $d(t)$ is still small ($F=0.1$, $0.2$, $0.4$), and it decreases when $d(t)$ approaches 
the value $1/4$ ($F=1.2$, $1.4$). Qualitatively, this is consistent with the quasi-equilibrium behavior: At low temperatures,  $U=3$ 
shows an insulating character, such that both the double occupancy and the dc-conductivity increase 
with temperature (Fig.~\ref{fig-quasiequi}c and d). At high temperatures, $\sigma_\text{dc}$ always {\em decreases} with $T$ 
($\sigma_\text{dc} \to 0$  for $T\to\infty$), while $d$ increases to the uncorrelated value $d=1/4$. Quantitatively, the 
current $j(t)$ in the driven system at $U=3$ remains below the quasi-equilibrium value, in particular for large $F$, 
where the energy increase is faster than the thermalization rate. 

In summary, the above discussion confirms that the existence of a quasi-steady current is tied to the lack of thermalization at 
large $U\gg\hopp$. Since the thermalization time depends exponentially on $U/\hopp$, the loss of the quasi-steady current 
occurs quite abruptly when $U$ is decreased. (A different regime of fast thermalization to infinite temperature has been 
encountered at $U=F$, when doublon-hole production requires only a single hopping process \cite{Mikelsons2012}.) 
That there is no steady current at smaller $U$ also explains why we could so far not really determine $\fth$ down to the metal-insulator transition. 
From now on, we will focus on regimes of large $U$ where the steady current does exist.


%
\begin{figure}[t]
\centerline{\includegraphics[width=16cm,clip=true]{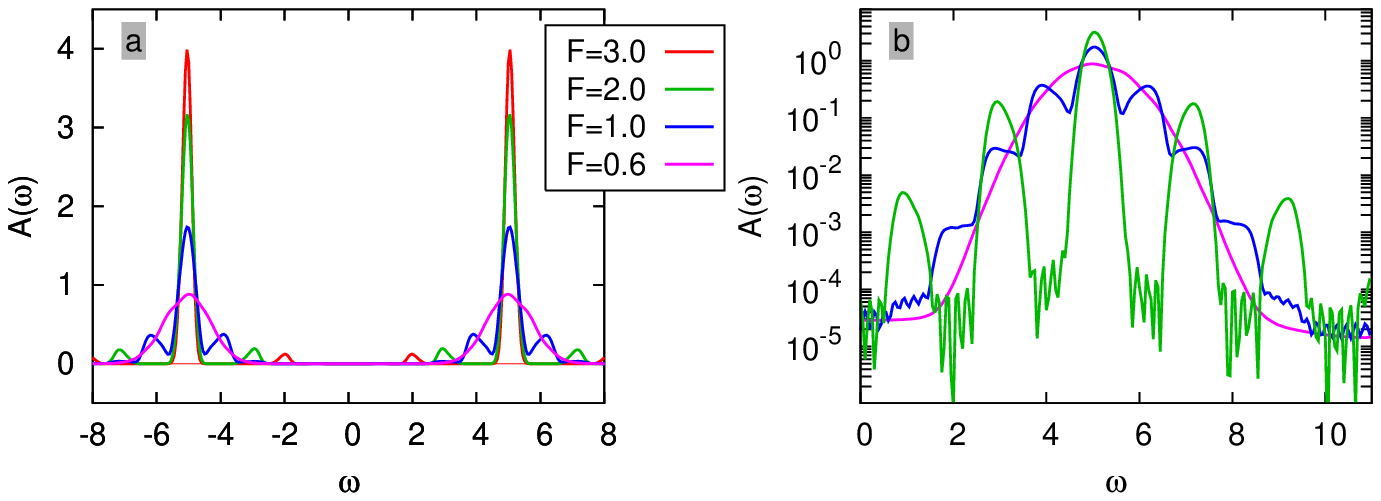}}
\centerline{\includegraphics[width=16cm,clip=true]{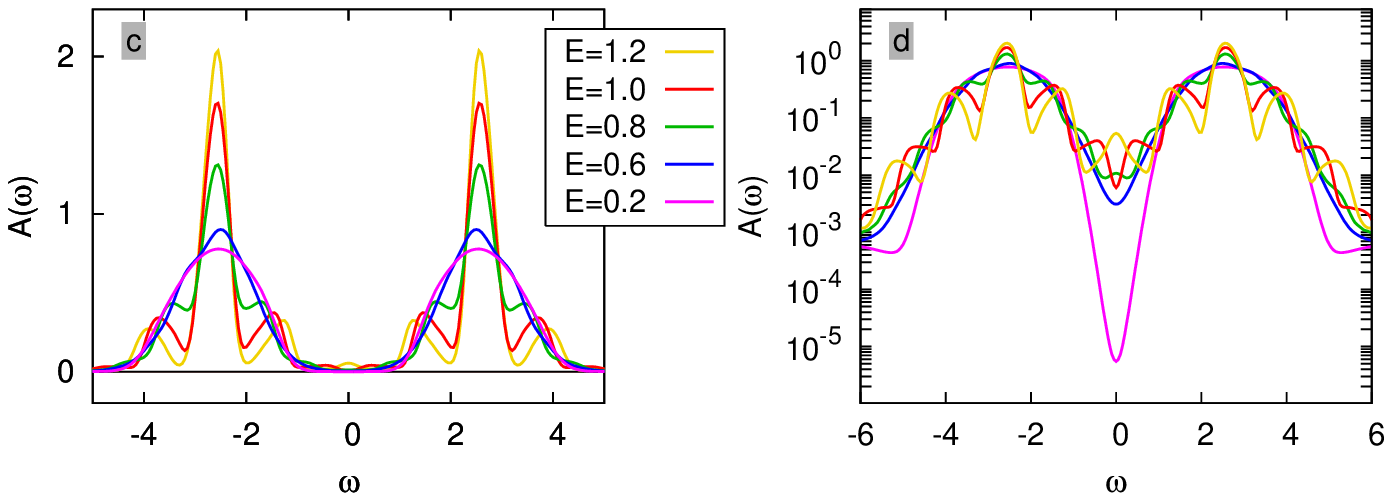}}
\caption{\label{fig-spectrum}
(a) Spectrum $A(\omega,t)$ for $t=10$, $U=10$, $\beta=5$, and various fields. (b) Same data on a logarithmic 
scale. The low-amplitude noise is due to a finite cutoff $s_\text{max}=40$ of the Fourier integral Eq.~(\ref{aret}).
(c) and (d)  Spectrum $A(\omega,t)$ for $t=10$, $U=5$, $\beta=5$. }
\end{figure}

\section{Spectral function}
\label{sec-spectrum}
\vspace{2mm}

Further insight into the quasi-steady current-carrying state can be gained from the spectral function,
Eq.~(\ref{aret}). In Fig.~\ref{fig-spectrum}a and b, we plot $A(\omega,t)$ for $U=10$. The values of $F$ 
in Fig.~\ref{fig-spectrum}a and b  are well below $\fth$ (which is large for $U=10$), such that there is essentially 
no current after the initial polarization of the system, and no change of either energy or double occupancy 
within the numerical accuracy. The state is thus truly stationary for all practical purposes, and $A(\omega,t)$ 
has no measurable $t$-dependence for $t\gtrsim10$. The spectrum consists of upper and lower Hubbard bands, 
separated by $U$. For large $F$, these Hubbard bands split into isolated peaks with a spacing $\Delta \omega =F$, 
which are narrowed with respect to the original Hubbard bands. 

The side-bands resemble the Wannier-Stark ladder for an electron in a tight-binding band with applied homogeneous 
field \cite{Wannier,Davies1988}. However, while all single-particle eigenstates in a tight-binding model with applied field 
are rigorously localized, we note that this is not true for the many-body situation of a doublon in the Mott insulator. A 
single doublon or hole which is added to the Mott insulator remains mobile even at large $F$, because its potential energy 
can be transferred to other particles or spin-excitations \cite{Mierzejewski2011a}. Nevertheless, the Hubbard side bands 
still reflect the localization which remains on short times, i.e., the side-peak at $\omega=U-nF$ may be interpreted as 
adding an electron at site $i$ into a many-body Wannier-Stark ``resonance'' that is localized at a different site $j$, with 
$({\bm R}_j-{\bm R}_i){\bm F}/|{\bm F}| = ({\bm R}_j-{\bm R}_i){\bm \eta}=n$.

At small fields, Wannier Stark side-bands are not clearly resolved, and we observe only a broadening of the spectrum
($F \lesssim 0.8$ in Fig.~\ref{fig-spectrum}b and d). At smaller $U$, this field-induced broadening eventually leads to 
a filling-in of the gap (Fig.~\ref{fig-spectrum}c and d), which indicates the possibility of the system to create a 
doublon-hole pair at no cost of energy. The Mott insulating ground state becomes unstable in the field. As soon as
the Wannier-Stark side-bands split off, $A(\omega=0)$ behaves non-monotonously as a function of $F$. This goes 
along with a non-monotonous $j$-$F$ curve, and explains the failure of Eq.~(\ref{fth}) at large $F$. At large 
$U$, where the system can tolerate strong enough fields such that the side-bands are well resolved, the 
non-monotonous behavior of the current even reveals clear resonances at integer $U/F$  (Fig.~\ref{fig-fth}c), 
although in this regime the steady current is superimposed with more and more irregular and longer lived 
transient oscillations. We note that the case $U/F=1$ is particularly interesting because the degenerate manifold 
of states is then described by an effective spin model with a quantum phase transition \cite{Sachdev2002}, 
but this physics shall not be addressed here.


%
\begin{figure}[t]
\centerline{\includegraphics[width=16cm]{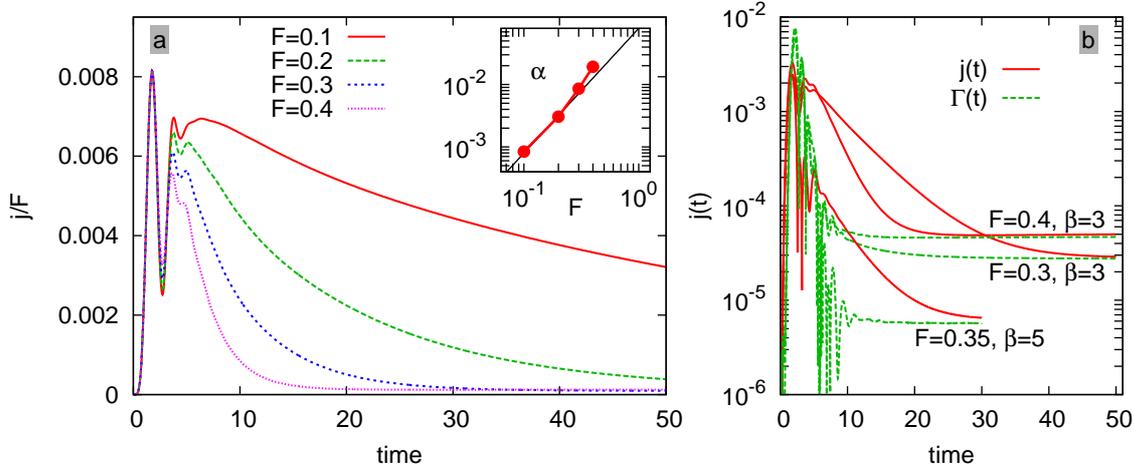}}
\caption{
\label{fig-highT}
(a)
Current for $U=4.5$ and high temperature $\beta=3$.
{\it Inset:}
The curves of the main plot are fit with an exponential 
$j(t)=A\exp(-\alpha t)+B$, and the rate $\alpha$ is plotted as a function of $F$.
The solid line corresponds to the relation $\alpha \propto F^{2}$.
(b)
Current $j(t)$ (solid line) and doublon creation current $\jd(t)$ (dashed line)
on a logarithmic scale, for $U=4.5$ and indicated values of $\beta$ and $F$.}
\end{figure}

\section{Influence of temperature}
\label{sec-temp}
\vspace{2mm}

We now turn to one of the most important discussions of this paper, and ask why doublons and holes that are created by 
field-induced tunneling do not lead to an increase of the current with time. Another question to be addressed here is the 
influence of the initial temperature on the dielectric breakdown, which actually turns out to be closely related: For a
hot initial state one has thermally excited carriers in addition to those which are induced by the field, and yet we will 
find that they do not influence the current at long times. 
In Fig.~\ref{fig-highT}a we demonstrate this behavior for $U=4.5$ and a rather high temperature, $\beta=3$. 
For these parameters, the linear response current $F\sigma_\text{dc}$ is actually larger than the field-induced 
current that is expected from Eq.~(\ref{fth}). However, $j(t)$ is found to be close to $F\sigma_\text{dc}$ only at 
early times, while it decreases almost to zero later on. This behavior was already described in {\prl}, where we also 
found that the final value is close to the zero-temperature field-induced current, although the decay to that value is 
very slow for small $F$, which hindered  a precise determination of the long-time limit for most parameters.

Because we are now using NCA, it is possible to take a closer look at the long-time behavior. To get a deeper understanding, 
it is useful to separately consider the component of the current that is related to the increase of 
the interaction energy, $(\frac{d}{dt} E_\text{int})/F$.  This turns out to be just the doublon-creation current that 
was defined ad-hoc in Eq.~(\ref{doublon-current}), $ (\frac{d}{dt} E_\text{int})/F = (1/F)\frac{d}{dt} U d(t) = \jd(t)$. 
Remarkably, we find that the long time limit of  $j(t)$ is very close to $\jd$, which by 
itself does not change much over the whole time interval (Fig.~\ref{fig-highT}b). This shows that  the current at 
long times is essentially determined by the doublon production rate, while the produced doublons themselves 
seem not to contribute to the current. This fact is also seen in Fig.~\ref{fig-current_U5b5}c, where we plot the
long-time average $\expval{\jd}_t$ as a function of $F$ in addition to the time average $\expval {j}_t$. 

To explain this behavior, one may wonder in the first place how charge carriers in correlated 
systems behave in the presence of large fields. This is well studied, e.g., in the $tJ$ model, where all terms changing
the doublon or hole numbers have been projected out, and in a strong field one is left with the motion
of quasiparticles. While noninteracting particles in a tight binding model are completely localized and 
perform Bloch oscillations (see, e.g., Ref.~\cite{Glueck2002}), the motion of carriers in a many-body system is 
a competition between Wannier-Stark localization and transfer of potential energy to other 
degrees of freedom, such as spins in the $tJ$ model  \cite{Vidmar2009,Mierzejewski2011a} 
or phonons in the Holstein model  \cite{Vidmar2011,Vidmar2011b}. Even in a large field, carriers 
remain mobile (the conductivity typically decreases with a power of $F$), and the decay of the 
linear response current in Fig.~\ref{fig-highT}a cannot be related to Wannier-Stark localization. 

On the other hand, the conductivity will always vanish at infinite 
temperature, because then potential energy can never be passed to other particles, spins, phonons 
etc.  This motivates the explanation that for a low doublon production rate, the system can 
essentially establish a state of infinite temperature for the kinetic energy of doublons and holes,
while their number is fixed (in contrast, infinite temperature for the interaction energy would imply $d=0.25$ 
as in Sec.~\ref{sec-quasi-failure}). More rigorously, to describe the short-time behavior one may project 
out all terms from the Hamiltonian that change the double occupancy (both field induced and
interaction induced). The resulting model $H_{\dh}$ is a generalized $tJ$ model, containing both 
doublons and holes in an external field $F$. Within this description, doublons and holes 
might then show quasi-equilibrium  behavior with a conductivity $\sigma_{\dh}$, and reach 
infinite doublon temperature $T_{\dh}$ much faster than the change of the doublon number happens 
via tunneling. For $T_{\dh}\to\infty$ one again has $\sigma_{\dh}=0$,  and the only remaining current is 
the $\dh$-creation current.

\begin{figure}[t]
\centerline{\includegraphics[width=16cm]{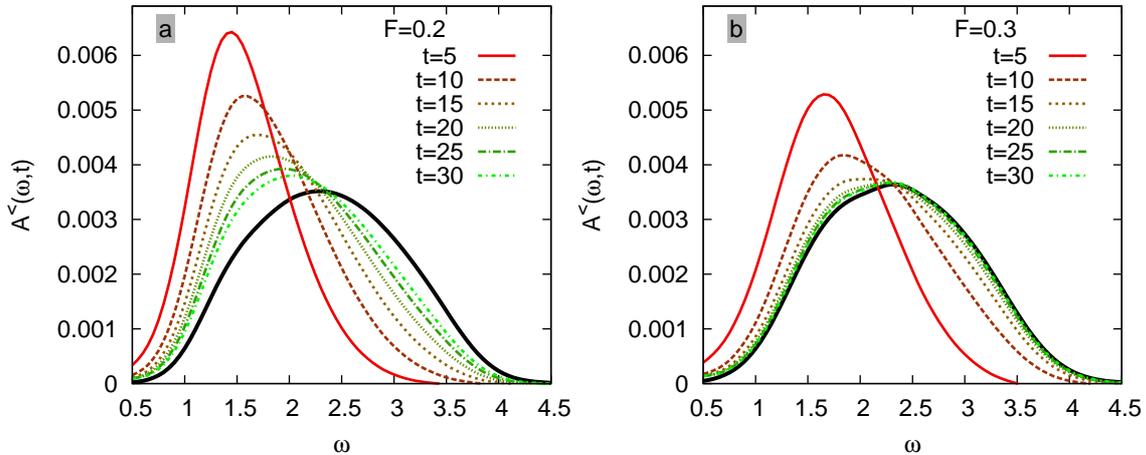}}
\caption{\label{fig-occ}
Occupation function $A^<(\omega,t)$ for $U=4.5$, $\beta=3$ in the upper Hubbard band, 
for various times. $F=0.2$ for (a), and $F=0.3$ for (b). The curves are compared to the spectral 
function $A(\omega,t)$ (rescaled by $0.0045$), which is almost time-independent for those 
parameters (bold black line).}
\end{figure}

Although we do not explicitly derive $H_{\dh}$ (we could not solve this model anyhow), there are two predictions 
of this simple argument which one can directly verify: (i) Quite generally, from a high-temperature expansion we expect that when $T\to\infty$ 
one has both $E_\text{tot}\sim1/T$
and $\sigma_\text{dc}\sim1/T$. In a quasi-equilibrium description, $j(t)$ should thus decay exponentially 
with a rate $\propto F^{-2}$ for $F\to0$, due to Eq.~(\ref{energy-increase}) \cite{Mierzejewski2010,Eckstein2011b}.  
This behavior is 
indeed
seen in the inset of Fig. \ref{fig-highT}a. 
(ii) To directly observe 
the ``heating'' of doublons and holes one can look at the occupation function, or
photoemission spectrum
Eq.~(\ref{ales}), which is plotted in Fig.~\ref{fig-occ} for $U=4.5$ and various times. The fields, $F=0.2$ and $0.3$, 
are chosen such that the total weight of $A^<(\omega,t)$ in the upper Hubbard band, which is roughly 
related to the total number of doublons, does not change much with time. The shape of $A^<(\omega,t)$,
however, changes considerably. In the initial state and shortly after the switch-on of the field, the
occupation is concentrated at low frequencies,
as expected for a thermal state with $A^<(\omega)=A(\omega)f(\omega)$. As time increases, 
$A^<(\omega,t)$
approaches a curve that is proportional to the spectral function itself (bold lines in Fig.~\ref{fig-occ}), 
corresponding to a flat distribution, $N(\omega)=const.$, characteristic of a state with zero kinetic 
energy (high-temperature state).


%
\begin{figure}[t]
\centerline{\includegraphics[width=16cm]{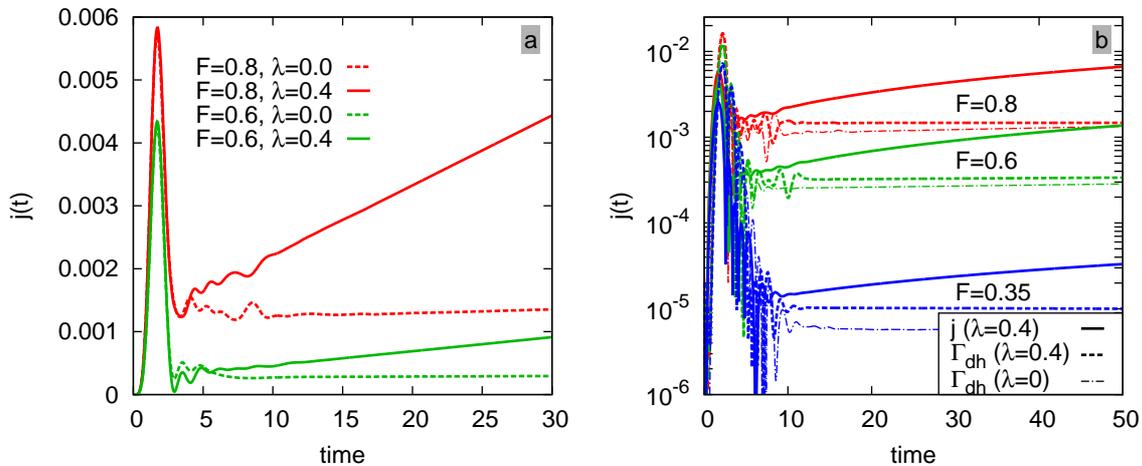}}
\caption{\label{fig-lambda}
(a) Comparison of the current at $U=4.5$, $\beta=10$ with dissipation (solid lines) and without dissipation 
(dashed lines).
(b) Comparison of the doublon-creation current (dashed lines) and the full electric current $j$ (solid lines) 
for $U=4.5$, $\beta=10$, and various fields $F$
($\lambda=0.4$).
}
\end{figure}

\section{Influence of dissipation}
\label{sec-diss}
\vspace{2mm}

In a condensed matter system, dissipation of energy to lattice-, spin-, and other degrees of freedom is unavoidable, and can 
happen on sub-picosecond timescales. It is thus essential to understand which of the findings of the previous section will
be robust against such processes. The steady states of dissipative systems with dc driving is itself an interesting area of 
research \cite{Amaricci2011,Han2012}. Within DMFT, dissipation terms can be included phenomenologically, either by coupling 
a local fermionic bath \cite{Tsuji2009, Amaricci2011,Aron2012} or a bosonic (``phonon'') bath \cite{Eckstein2012c} at fixed 
temperature. Here we use the latter approach, since it guarantees particle number conservation and dissipates only energy. 

Following Ref.~\cite{Eckstein2012c}, the electronic self-energy in this case is supplemented by a bath contribution, 
which is the lowest order diagram for a Holstein-type electron-phonon coupling,  $\Sigma_\text{diss}[G]=\lambda^2 
G(t,t') D(t,t')$. here, $\lambda$ measures the coupling strength, and $D(t,t')$ is the equilibrium propagator for a boson 
with energy $\omega_0$; $D(t,t')=-i \text{Tr} [ \text{T}_\mathcal{C} \exp(-i \int_\mathcal{C} dt \omega_0 b^\dagger b) 
b(t)b^\dagger(t')]/Z $. The temperature $1/\beta$ of the bath is fixed, such that the bath has no memory ($\omega_0=\hopp$ 
in the following). It is important to note that we choose $\omega_0 = \hopp \ll U $  for the phonon frequency, to prevent 
opening a new channel for doublon-hole recombination via phonon emission. In general, the precise way of including 
the dissipation should not matter too much, and we chose parameters such that the equilibrium physics remains 
almost unchanged by the presence of dissipation. 
(In contrast, for a fermionic bath in the wide-band limit the equilibrium state would be modified due to mid-gap states 
that are created in the spectrum of the insulator with dissipation \cite{Aron2012}.)

Figure \ref{fig-lambda}a compares the current $j$ at $U=4.5$, with and without dissipation. In contrast to the isolated 
system ($\lambda=0$), the current in a system with dissipation never becomes stationary, but it increases more or 
less linearly with time ($\lambda=0.4$). Based on the results of the previous sections, the explanation is straightforward: 
If the system can dissipate energy, doublons and holes will never reach $T_{\dh}=\infty$, and the distribution $N(\omega,t)$ 
never becomes flat. Hence doublons and holes contribute to the current at all times. Since their number is increasing due 
to the field-induced doublon-hole production, the current increases with time. The doublon-creation current $\jd$, on the 
other hand, is not expected to depend strongly on the coupling to phonons with an energy $\omega_0 \ll U$. This is in fact true, as 
apparent from Fig.~\ref{fig-lambda}b, where we compare the currents $j$ and $\jd$ for dissipative systems with 
electron-phonon coupling $\lambda=0.4$. The qualitative behavior of $\jd$ is the same with and without dissipation, i.e., 
the value is time-independent and increases exponentially with $F$.

\section{Implication for experiments}
\label{experiments}
\vspace{2mm}

\begin{figure}[t]
\centerline{\includegraphics[width=16cm]{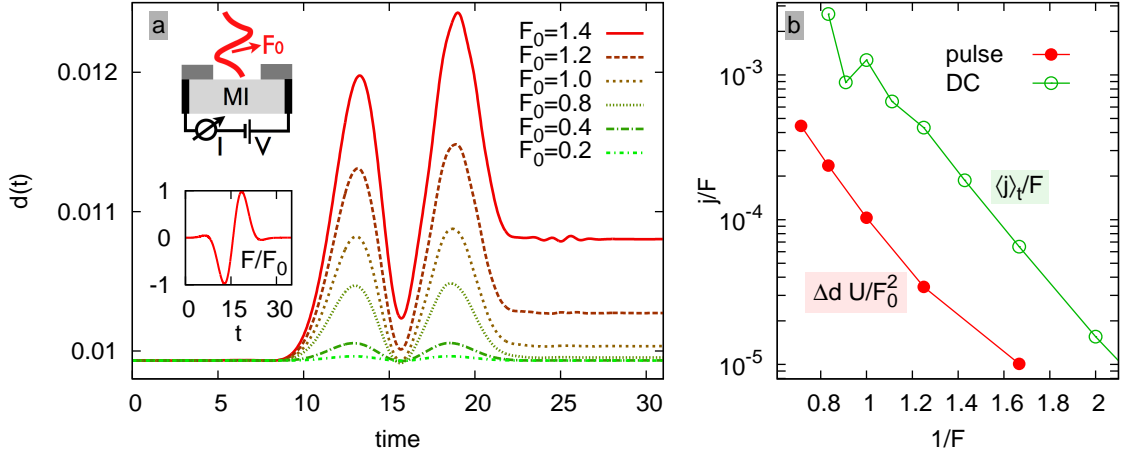}}
\caption{\label{fig-thz}
(a) $d(t)$ during a single-cycle THz-pulse with amplitude $F_0$,
for $U=5$ and $\beta=5$.
{\it Inset:} The electric field pulse, Eq.~(\ref{thzpulse}), with $\tau=3.9269$.
(b) Integrated doublon current, $ \expval{\jd}_\text{av}/F_0 = U \Delta d / F_0^2$ (filled symbols), where $\Delta d$ 
is the change of the double occupancy over the field cycle in panel (a). The same parametrization as in Fig.~\ref{fig-fth}a is used. 
For comparison, we include the dc curve $\expval{j}_t/F$ for $U=5$, taken from Fig.~\ref{fig-fth}a  (open symbols). 
Both have the same slope, related to $\fth$.}
\end{figure}

To design an experiment that can potentially measure the threshold behavior, one should account for the coupling of 
the sample to leads \cite{Okamoto2008a,Meisner2010}. Doublons and holes can then escape through the boundary, 
such that their density in the sample becomes stationary and a true steady state is reached. (Otherwise, the 
quasi-steady state would be quite short-lived: extrapolating the linear increase of $d$ in Fig.~\ref{fig-current_U5b5}b, 
e.g., gives a time of a few 1000 until the uncorrelated value $d=0.25$ would be reached, corresponding to a few picoseconds for 
$\hopp=1eV$.) Nevertheless, in the dc setup heating of the system is quite substantial, and the current will be 
a balance between non-perturbative doublon-hole production and dissipative effects. 
To avoid these complications, one could probe the strong field behavior by using short pulses, as already 
proposed for dielectrics (band insulators) \cite{Apalkov2012}. If the pulse frequency $\Omega \ll U$ is below the 
Mott gap, e.g., in the THz range, the doublon production might be described by the dc results to a good approximation.
If a weak external bias $V$ is applied to the sample in addition to the strong
THz pulse, the induced charge is collected at the leads. The current, averaged over many THz pulses,
would then be proportional to the total number $\Delta d$ of doublons created per pulse. 
In Fig.~\ref{fig-thz}, we have simulated the outcome of such an experiment (as sketched in the inset). 
Figure~\ref{fig-thz}a shows the time-evolution of the double occupancy during a single-cycle pulse
\begin{equation}
\label{thzpulse}
F(t)=F_0 \sin( \Omega (t-t_0)) \, e^{-\frac{1}{2}\left(\frac{t-t_0}{\tau}\right)^2}, 
\end{equation} 
centered around $t_0=2\pi/\Omega$, with frequency $\Omega=0.4$, and a Gaussian envelope 
(see inset in Fig.~\ref{fig-thz}a). Motivated by our results, we define an averaged doublon creation 
current $\expval{\jd}_\text{av} = U/F_0\,\Delta d$. As seen in Fig.~\ref{fig-thz}b, the threshold 
behavior of $\expval{\jd}_\text{av}$ turns out to be the same as in the dc case (Fig.~\ref{fig-fth}).
This seems reasonable if the doublon production is determined by the largest field during the cycle.

Using time-resolved photoemission spectroscopy, one might be able to observe the heating effect demonstrated 
in Fig.~\ref{fig-occ}: In a system with strong external bias, but not strong enough to lead to a dielectric breakdown, 
one may suddenly excite carriers created by photo-doping, monitor the evolution of the distribution 
function, and compare it to the result without applied field.

\section{Summary and Conclusion}
\label{discussion}
\vspace{2mm}

In conclusion, we have used DMFT with the NCA-based impurity solver to study the behavior of a Mott insulator in 
strong dc electric fields $F$. The electric current $j$ is given by a contribution $\jd$ which can be  associated with the 
doublon-hole production due to field-induced tunneling [Eq.~(\ref{doublon-current})], and a contribution related to the 
conductance of doublons and holes which are either thermally excited or induced by the field. For an isolated 
system, however, these carriers accumulate 
energy from the field and rapidly reach an ``infinite 
temperature'' state with zero conductivity, as apparent from a flat occupied density of states. This explains the 
peculiar steady state that has been described earlier, in which the current is time-independent (and given by $\jd$) 
although the number of carriers constantly increases. In an isolated system, the current itself is thus a good 
measure for the field-induced tunneling. In contrast, when dissipation of energy is taken into account, carriers 
cannot reach infinite temperature. The current now {\em does} depend on the number of doublons, temperature,
and the coupling to the environment, 
whereas the doublon-hole creation current $\jd$ still provides an intrinsic measure 
of the tunneling. For fields $F \lesssim \hopp$, $\jd$ increases with $F$ with a  threshold behavior, Eq.~(\ref{fth}). 
For larger fields,  we find that the $j$-$F$ characteristics reveals peaks at integer values of $U/F$, which are 
related to signatures of Wannier-Stark localization of carriers in the spectrum.

The threshold field might be determined experimentally by measuring the current induced by a field pulse under weak bias, 
while the heating of induced doublons should be observable with time-resolved photoemission. 

\ack
We thank T. Oka  for useful discussions. PW acknowledges support from FP7/ERC starting grant No. 278023.

\section*{References}
\newcommand{\mylink}[1]{}
\newcommand{\mytitle}[1]{}

\end{document}